\documentclass[aps,prl,twocolumn,superscriptaddress,nopacs,citeautoscript]{revtex4}

\usepackage{graphicx}
\usepackage{amsfonts}
\usepackage{amssymb}
\usepackage{color}

\newcommand\ea{\textit{et al.}}

\begin{document}


\title{Structural origin of apparent Fermi surface pockets in angle-resolved photoemission of Bi$_2$Sr$_{2-x}$La$_x$CuO$_{6+\delta}$}


\author{P.~D.~C.~King}
\affiliation{School of Physics and Astronomy, University of St Andrews, North Haugh, St Andrews, KY16 9SS, United Kingdom}

\author{J.~A.~Rosen}
\affiliation{Department of Physics and Astronomy, University of British Columbia, Vancouver, British Columbia V6T 1Z1, Canada}

\author{W.~Meevasana}
\affiliation{School of Physics and Astronomy, University of St Andrews, North Haugh, St Andrews, KY16 9SS, United Kingdom}
\affiliation{School of Physics, Suranaree University of Technology, Nakhon Ratchasima, 30000 Thailand}

\author{A.~Tamai}
\author{E.~Rozbicki}
\affiliation{School of Physics and Astronomy, University of St Andrews, North Haugh, St Andrews, KY16 9SS, United Kingdom}

\author{R.~Comin}
\author{G.~Levy}
\author{D.~Fournier}
\affiliation{Department of Physics and Astronomy, University of British Columbia, Vancouver, British Columbia V6T 1Z1, Canada}

\author{Y.~Yoshida}
\author{H.~Eisaki}
\affiliation{National Institute of Advanced Industrial Science and Technology, Tsukuba, Ibaraki 305-8568, Japan}

\author{K.~M.~Shen}
\affiliation{Department of Physics, Cornell University, Ithaca, New York 14853, USA}

\author{N.~J.~C.~Ingle}
\affiliation{Advanced Materials and Process Engineering Laboratory, University of British Columbia, Vancouver, British Columbia V6T 1Z1, Canada}
\author{A.~Damascelli}
\affiliation{Department of Physics and Astronomy, University of British Columbia, Vancouver, British Columbia V6T 1Z1, Canada}
\affiliation{Quantum Matter Institute, University of British Columbia, Vancouver, British
Columbia V6T 1Z4, Canada}

\author{F.~Baumberger}
\email{fb40@st-andrews.ac.uk}
\affiliation{School of Physics and Astronomy, University of St Andrews, North Haugh, St Andrews, KY16 9SS, United Kingdom}

\begin{abstract}
We observe \emph{apparent} hole pockets in the Fermi surfaces of single-layer Bi-based cuprate superconductors from angle-resolved photoemission (ARPES). 
From detailed low-energy electron diffraction measurements and an analysis of the ARPES polarization-dependence, we show that these pockets are not intrinsic, but arise from multiple overlapping superstructure replicas of the main and shadow bands. 
We further demonstrate that the hole pockets reported recently from ARPES [Meng~\ea, Nature {\bf 462}, 335 (2009)] have a similar structural origin, and are inconsistent with an intrinsic hole pocket associated with the electronic structure of a doped CuO$_2$ plane. The nature of the Fermi surface topology in the enigmatic pseudogap phase therefore remains an open question.\end{abstract}

\maketitle

\begin{figure*}
\includegraphics[width=0.8\textwidth]{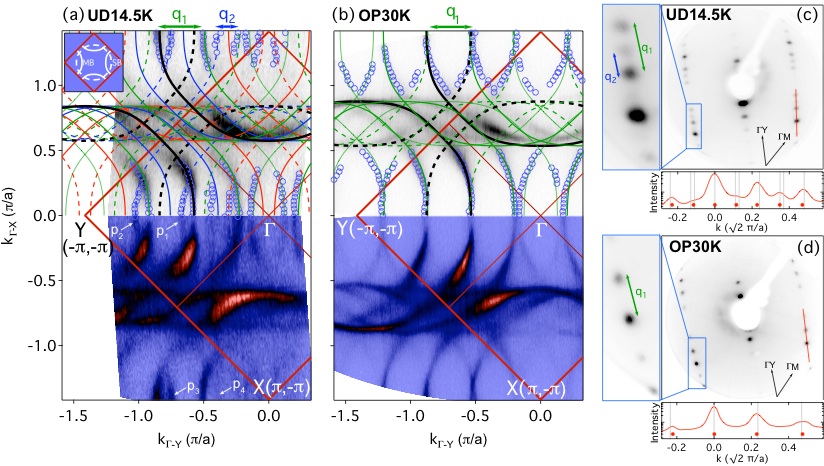}
\caption{\label{f:FSM_LEED}(color online) ARPES Fermi surface of (a) UD14.5K and (b) OP30K La-Bi2201, measured using $p$-polarization. Extracted contours (blue circles) and a tight-binding model of the main (solid), shadow (dashed), and $\pm\mathbf{q}_1$ or $\pm2\mathbf{q}_1$ (green), $\pm\mathbf{q}_2$ (blue), and $\pm(\mathbf{q}_1+\mathbf{q}_2)$ (red) Umklapp bands are overlaid on the data. Overlapping bands due to the multiple superstructures describe the apparent Fermi pockets $p_1$ -- $p_4$. The superstructure vectors of $q_1=0.245\pm0.015$ and $q_2=0.130\pm0.015$ for UD14.5K, and $q_1=0.235\pm0.015$ for OP30K La-Bi2201, are taken from an analysis of LEED, shown in (c) and (d), respectively. A magnified view of the blue region and a line cut along the $\Gamma$-Y direction (red line) are shown to the left of and below each pattern, respectively. Extracted peak positions (red dots) and those expected for superstrucure peaks at $\mathbf{k}_0\pm{m_1}\mathbf{q}_1\pm{m_2}\mathbf{q}_2$ (vertical grey lines) are in good agreement. The inset to (a) shows a simplified Fermi surface with only the main (MB) and shadow (SB) bands.}
\end{figure*} 

The pseudogap is one of the defining properties of the hole-doped high-T$_c$ superconductors~\cite{Timusk:Rep.Prog.Phys.:62(1999)61--122}. Understanding its origin is widely regarded as a key to unravelling the mechanisms of the high transition temperature superconductivity in these materials. While above the pseudogap temperature, T$^*$, a large closed Fermi surface characteristic of ordinary metals is observed~\cite{Kanigel:NaturePhys.:2(2006)447--451,Hashimoto:NatPhys:6(2010)414--418}, at temperatures below T$^*$, angle-resolved photoemission (ARPES) measurements instead revealed disconnected Fermi ``arcs'', centered around the $(0,0)\to(\pi,\pi)$ nodal direction~\cite{Norman:Nature:392(1998)157--160,Shen:Science:307(2005)901--904,Kanigel:NaturePhys.:2(2006)447--451,Valla:Science:314(2006)1914--1916}. It remains unclear how the existence of such open-ended sections of Fermi surface can be reconciled with the recent observation of quantum oscillations in underdoped cuprates, indicative of small closed pockets of the Fermi surface in high field~\cite{Chakravarty:Phys.Rev.B:68(2003)100504,Yang:Phys.Rev.B:73(2006)174501,Doiron-Leyraud:Nature:447(2007)565--568,LeBoeuf:Nature:450(2007)533--536,Yang:Nature:456(2008)77--80,Yelland:Phys.Rev.Lett.:100(2008)047003,Sebastian:Nature:454(2008)200--203,Meng:Nature:462(2009)335--338,Yang:arXiv:1008.3121:(2010),Norman:Physics:3(2010)86}. 

Meng~\ea~\cite{Meng:Nature:462(2009)335--338} recently reported measurements of the Bi-based cuprates Bi$_2$Sr$_{2-x}$La$_x$CuO$_{6+\delta}$ (La--Bi2201) which appeared to show, for the first time, the existence of closed Fermi surface pockets in the pseudogap phase from ARPES. However, Bi-based cuprates are structurally highly complex. They possess an orthorhombic lattice distortion~\cite{Subramanian:Science:239(1988)1015--1017,Torardi:Phys.Rev.B:38(1988)225--231}, which is likely the origin of the so-called shadow Fermi surface~\cite{Mans:Phys.Rev.Lett.:96(2006)107007,Nakayama:Phys.Rev.B:74(2006)054505,Aebi:Phys.Rev.Lett.:72(1994)2757--2760}. In addition, they are prone to single~\cite{Subramanian:Science:239(1988)1015--1017,Torardi:Phys.Rev.B:38(1988)225--231,Withers:J.Phys.C:21(1988)L417,Withers:J.Phys.C:21(1988)6067} or even multiple~\cite{Yang:PhysicaC:308(1998)294--300} superstructure modulations along the crystallographic $b$-axis, caused by a slight lattice mismatch between the BiO and CuO$_2$ plane. The separation of subtle electronic effects intrinsic to the CuO$_2$ plane from the diffraction replica (DR) of electronic bands resulting from structural complications is thus often non-trivial~\footnote{Throughout, we distinguish the $(\pi,\pi)$ Umklapp of the main band, giving rise to the shadow band, from superstructure diffraction replica to be consistent with conventional terminology.}. 

Here, we show that large superstructure periodicities of up to 14~a$_0$, coexisting with the already well-established periodicity of $\approx4.2$~a$_0$, are common in Bi2201 samples, such as those investigated by Meng \textit{et al.}~\cite{Meng:Nature:462(2009)335--338}. The same multiple periodicities can be observed in the DR of electronic bands and are even observed following Pb-doping, which is known to suppress superstructure effects in Bi-based cuprates. The resulting presence of multiple, and overlapping, bands in ARPES measurements leads to imitations of closed Fermi-surface pockets. Analyzing their polarization dependence, we demonstrate that the front and back sides of these pockets derive from the main and shadow bands, respectively. Thus, the pockets observed here and by Meng~\ea~\cite{Meng:Nature:462(2009)335--338} reflect structural artifacts, rather than the intrinsic electronic structure of the doped CuO$_2$ plane.

We investigated optimally- (OP) and under- (UD) doped La-Bi2201 with $x=0.5$ (OP30K), $x=0.75$ (UD$20$K), and $x=0.8$ (UD14.5K), and optimally-doped Bi$_{1.7}$Pb$_{0.35}$La$_{0.4}$Sr$_{1.6}$CuO$_{6+\delta}$ [(Pb,La)-Bi2201, OP35K] samples. 
ARPES measurements on La-Bi2201 (Figs. 1,3) were performed in the pseudogap phase at $\sim\!\!17.5$~K and $\sim\!\!33.5$~K for UD14.5K and OP30K La-Bi2201, respectively, using linearly-polarized He-I$\alpha$ radiation ($h\nu=21.22$~eV) and a SPECS Phoibos 225 hemispherical analyzer, while the data on (Pb,La)-Bi2201 (Fig. 2) was taken in the superconducting phase at 10~K with unpolarized light.
The angular and energy resolutions for all measurements were set to $0.3$$^\circ$ and better than $20$~meV, respectively.
All LEED patterns shown here were recorded at temperatures between $\sim$T$_c$ and $\sim\!\!2$T$_c$ using an incident electron energy of 35~eV.

Fig.~\ref{f:FSM_LEED}(a) shows the Fermi surface of UD14.5K La-Bi2201 as measured by ARPES. Fermi arcs are visible centred around the nodal directions, although their intensity is suppressed along $\Gamma$--Y due to matrix element effects which we shall discuss later. Multiple copies of these arcs can be observed, separated by $\sim\!0.28$~\AA$^{-1}$ along the $\Gamma$--Y direction, consistent with DR corresponding to the established dominant superstructure periodicity of this material. In addition, weak features with opposite dispersion to the Fermi arcs appear to form several small closed pockets along the nodal direction (white arrows, $p_1$ -- $p_4$). These features cannot be observed in optimally-doped La-Bi2201, shown in Fig.~\ref{f:FSM_LEED}(b), consistent with the findings of Meng~\ea~\cite{Meng:Nature:462(2009)335--338}.

In the following, we show that the closed pockets appear as a natural consequence of structural complications in La-Bi2201. Lines of periodic diffraction maxima, characteristic of superstructure modulation along the $\Gamma$--Y direction, are clearly discernible in the LEED pattern from OP30K La-Bi2201 (Fig.~\ref{f:FSM_LEED}(d)). From their spacing, the superstructure vector can be determined as $\mathbf{q}_i=(q_i,q_i)\frac{\pi}{a}$ with $q_1=0.235\pm0.015$, in agreement with the DR in the ARPES and with earlier diffraction studies~\cite{Dudy:J.Supercond.Nov.Magn.:22(2009)51--55}. 
Intriguingly, LEED from UD14.5K La-Bi2201 (Fig.~\ref{f:FSM_LEED}(c)) shows not only a similar superstructure vector $q_1=0.245\pm0.015$, but
exhibits yet further diffraction maxima revealing the co-existance of a second superstructure with $q_2=0.130\pm0.015$.
In order to demonstrate how these superstructure periodicities lead to the impression of hole-pockets in ARPES, we first fit a tight-binding model to the Fermi surface of the main band for each doping (solid black lines in Fig.~1(a),(b)), and then translate this band by $(\pi,\pi)$ to describe the shadow band resulting from the orthorhombic distortion. Finally, we add Umklapp bands, that is, DR of the main and shadow bands, using the $\mathbf{q_i}$ vectors determined independently from LEED. Without any further adjustment this simple model reproduces the entire measured Fermi surfaces for the UD14.5K and the OP30K samples over an extensive k-space range. In particular, it describes all apparent hole pockets in the underdoped sample and the absence of these pockets in optimally doped La-Bi2201.
For example, the pockets p$_1$ and p$_3$ in UD14.5K La-Bi2201 are created by the $\mathbf{q}_2$ DR of the shadow band crossing the main band, while another pocket (p$_2$) is formed by the $-\mathbf{q}_1$ DR of the main band overlapping the $-\mathbf{q}_2$ DR of the shadow band. All of these pockets are absent in OP30K La-Bi2201, which does not show the $\mathbf{q}_2$ periodicity in LEED.
It is therefore evident that the seemingly closed portions of Fermi surface in underdoped La-Bi2201 are not intrinsic, but appear from the overlapping of DR resulting from multiple superstructure periodicities.

\begin{figure}
\includegraphics[width=\columnwidth]{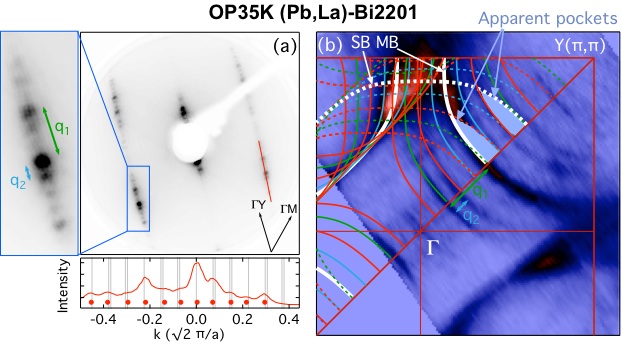}
\caption{\label{f:Pb}(color online) (a) LEED pattern and (b) ARPES Fermi surface of OP35K (Pb,La)-Bi2201, showing the presence of two superstructure vectors, and the resulting apparent pockets in the Fermi surface. A tight-binding model, using superstructure vectors determined from the LEED analysis, is also shown in (b), in good agreement with the data.} 
\end{figure}
We stress that the appearance of these pockets is not directly tied to the doping level (under- versus optimally-doped), or presence of a pseudogap, but rather the particular structural modulations. Indeed, we also observe multiple superstructures, with vectors of $q_1=0.225\pm0.015$ and $q_2=0.072\pm0.015$, in optimally-doped (Pb,La)-Bi2201 (Fig.~\ref{f:Pb}(a)). 
Although Pb-doping tends to suppress superstructure-related features in ARPES from Bi-based cuprates, the measured Fermi surface (Fig.~\ref{f:Pb}(b)) clearly shows the presence of Umklapps resulting from these superstructure vectors and their combinations, consistent with a tight-binding model using the $\mathbf{q_i}$ vectors determined from LEED. 
Similar to the situation in underdoped La-Bi2201, the overlapping of several of these bands gives rise to features that appear as closed Fermi surface pockets. However, as for UD14.5K La-Bi2201, this is entirely due to structural effects, and should not be confused with either an intrinsic hole pocket or an incommensurate density-wave order.

\begin{figure*}
\includegraphics[width=\textwidth]{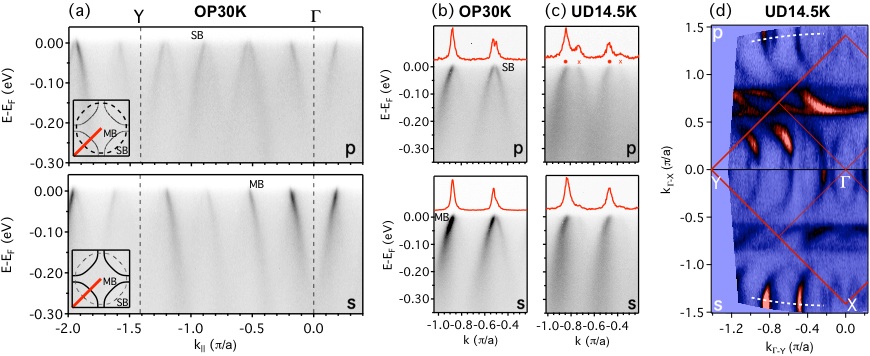}
\caption{\label{f:Pol}(color online) Polarization dependence of ARPES. (a) Dispersion along $\Gamma$--Y of OP30K La-Bi2201. (b),(c) Dispersion close to the equivalent direction in the 2$^{nd}$ Brillouin zone [along dashed lines in (d)] of OP30K and UD14.5K La-Bi2201. (d) ARPES Fermi surface of UD14.5K La-Bi2201. All spectra were measured using p (top) and s (bottom) polarization.}
\end{figure*}
The structural origin of the observed pockets can be further confirmed by polarization-dependent ARPES measurements, as shown in Fig.~\ref{f:Pol}. The ${{x^2}-{y^2}}$ symmetry of the hole in the Cu $d$-shell is odd with respect to the $\Gamma$--Y direction. Consequently, for the experimental geometry employed here, with the incident light and detected electrons both within the horizontal plane, the main band can be observed when measuring along the $\Gamma$--Y azimuth using $s$-polarized light, but is suppressed for $p$-polarization. This is the reason why the intensity along the Fermi arc diminishes approaching the $\Gamma$--Y nodal line in Fig.~\ref{f:FSM_LEED}(a,b). 
In contrast, the shadow band, which has the opposite parity of the main band~\cite{Mans:Phys.Rev.Lett.:96(2006)107007}, is visible in $p$-polarization, but not using $s$-polarization. This switching of intensities is clearly seen for OP30K La-Bi2201 in Fig.~\ref{f:Pol}(a)
\footnote{The lack of complete suppression of the forbidden transitions is due to the finite degree of polarization ($\sim\!80$\%) of the incident light in the experimental set-up used here.}. For measurements in the 2$^{nd}$ Brillouin zone, along the cut shown in Fig.~\ref{f:Pol}(d), the polarization is no longer strictly $s$ or $p$ since the sample is tilted off-vertical by $\sim\!30^\circ$.
Nevertheless, a strong relative intensity variation can still be observed between the main and shadow bands on switching from dominant $p$-polarization ($I_{MB}:I_{SB}$ smaller) to $s$-polarization ($I_{MB}:I_{SB}$ larger), as shown in Fig.~\ref{f:Pol}(b). 
The equivalent dispersion measured in UD14.5K La-Bi2201 is shown in Fig.~\ref{f:Pol}(c). Using $p$-polarization, two strong dispersions can be seen due to the main band and its $-\mathbf{q}_1$ DR, with two weaker neighbouring bands which form the back-side of the apparent Fermi surface pockets (marked in the Fermi level momentum distribution curve (MDC) by circles and crosses, respectively). This gives the appearance that the pocket on the Fermi surface is hole-like, as claimed in Ref.~\cite{Meng:Nature:462(2009)335--338}. However, on switching to $s$-polarization, these `pocket-forming' bands are strongly suppressed relative to the main bands, as is the spectral intensity of the back-side of all of the pocket features which can be seen in the ARPES Fermi surface (Fig.~\ref{f:Pol}(d)). This switching of intensities due to different parities of the front- and back--side of the FS pockets is difficult to reconcile with intrinsic pockets of a reconstructed Fermi surface and confirms that these features are derived from the shadow band.

\begin{figure}
\includegraphics[width=\columnwidth]{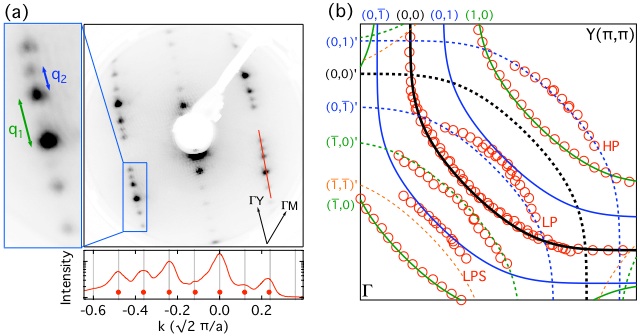}
\caption{\label{f:Meng}(color online) (a) LEED of UD$20$K La-Bi2201, very similar in composition to the UD18K sample considered by Meng~\ea~\cite{Meng:Nature:462(2009)335--338}, showing the presence of two superstructure vectors. (b) Red circles reproduce the Fermi surface contours of UD18K La-Bi2201 from Ref.~\cite{Meng:Nature:462(2009)335--338}. A tight-binding Fermi surface including Umkapp bands derived from our LEED analysis reproduces all features of the data from Meng~\ea~\cite{Meng:Nature:462(2009)335--338} including the apparent Fermi surface pockets (labelled LP, HP, and LPS after Ref.~\cite{Meng:Nature:462(2009)335--338}). The tight-binding bands are labelled by $(m_1,m_2)^{[']}$ where $m_i$ is the order of the $(\mathbf{q}_1,\mathbf{q}_2)$ superstructure replica, and a prime denotes the shadow band.}
\end{figure}
Given this, one must reconsider the analysis of Ref.~\cite{Meng:Nature:462(2009)335--338} and revise their conclusions regarding intrinsic hole pockets. Meng~\ea~\cite{Meng:Nature:462(2009)335--338} considered only Umklapp bands arising from a $q_1\approx0.24$ superstructure modulation, and found that these DR could not explain their data. However, LEED from La-Bi2201 with very similar composition to the UD18K sample of Ref.~\cite{Meng:Nature:462(2009)335--338} shows not only the $q_1=0.24$ superstructure, but also a second superstructure with $q_2=0.12\pm0.015$ (see Fig.~\ref{f:Meng}(a)). In Fig.~\ref{f:Meng}(b), we show that the measured pockets, and indeed all features of the Fermi surface topology extracted in Ref.~\cite{Meng:Nature:462(2009)335--338}, are accurately described by a tight-binding model of only the conventional main and shadow bands, provided DR are included corresponding to both of these superstructure vectors. 
Together with the polarization-dependent ARPES presented above, this 
shows unambiguously that the observations of Meng~\ea~\cite{Meng:Nature:462(2009)335--338} have a trivial interpretation, and cannot be taken as evidence for elusive Fermi pockets. 

We also note that a structural origin of the apparent hole pockets explains several puzzling observations of Ref.~\cite{Meng:Nature:462(2009)335--338}. 
First, the main band (Fermi arc) was observed to be longer than the back side of the claimed pocket (as is also evident here in Figs.~\ref{f:FSM_LEED}(a) and~\ref{f:Pb}(b). The lower intensity of the shadow band and its DRs, which appear to ``close'' the pockets, compared to the main band naturally accounts for this seemingly contradictory coexistence of Fermi arcs and hole pockets. 
Second, the spectral weight of the back side of the pockets in Ref.~\cite{Meng:Nature:462(2009)335--338} appears largest close to the nodal line. This is in contrast to theoretical expectations for an intrinsic pocket~\cite{Chakravarty:Phys.Rev.B:68(2003)100504,Yang:Phys.Rev.B:73(2006)174501}, but consistent with a superstructure replica of the shadow band. 
Third, multiple hole pockets were observed in Ref.~\cite{Meng:Nature:462(2009)335--338}, attributed to $q=0.24$ superstructure replica of a single intrinsic pocket. However, the front- to back-side spectral weight ratio differs for these pockets. Again this suggests that the two sides of the pockets derive from different bands, namely the main and shadow band, whose Umklapps display complex intensity variation due to matrix element effects.
Finally, we note that the superstructure vectors depend sensitively on doping as shown from the LEED analysis presented here, which provides a simple explanation for the unusual doping dependence of the apparent hole pockets reported by Meng~\ea~\cite{Meng:Nature:462(2009)335--338}.

In conclusion, we have shown that the impression of closed Fermi surface pockets can be created in the ARPES Fermi surface of Bi-based cuprates due to Umklapp bands arising from multiple incommensurate superstructures. The apparent Fermi surface pockets recently observed by Meng~\ea~\cite{Meng:Nature:462(2009)335--338} in La-Bi2201 are straightforwardly explained by such structural effects, and hence do not provide evidence for intrinsic hole pockets. We note that these findings do not exclude the presence of intrinsic Fermi surface pockets in cuprates, where the back side of the pocket has negligible spectral weight in ARPES measurements \cite{Yang:Nature:456(2008)77--80,Yang:arXiv:1008.3121:(2010)}. However, we remark that to date, clearly discernible hole pockets have only been reported from ARPES measurements in Bi-based and La-based cuprate systems~\cite{Meng:Nature:462(2009)335--338,Chang:NewJ.Phys.:10(2008)103016}, which are both subject to structural distortions. No such observation has been made in compounds free of such distortions, for example CCOC~\cite{Shen:Science:307(2005)901--904} and YBCO~\cite{Hossain:NaturePhys.:4(2008)527--531}.

\begin{acknowledgments}
We are grateful for useful discussions with Z.-X.~Shen, W.-S.~Lee, M.~Hashimoto, and I.~Vishik. The work in St Andrews was supported by the Scottish Funding Council, the European Research Council, and the UK EPSRC (Grant No.~EP/F006640). WM acknowledges The Thailand Research Fund for financial support. The work at UBC was supported by the Killam Program (A.D.), the A.\ P.\ Sloan Foundation (A.D.), the CRC Program (A.D.), NSERC, CFI, CIFAR Quantum Materials, and BCSI.

\end{acknowledgments}

\bibliographystyle{apsrev}

\begin{thebibliography}{29}
\expandafter\ifx\csname natexlab\endcsname\relax\def\natexlab#1{#1}\fi
\expandafter\ifx\csname bibnamefont\endcsname\relax
  \def\bibnamefont#1{#1}\fi
\expandafter\ifx\csname bibfnamefont\endcsname\relax
  \def\bibfnamefont#1{#1}\fi
\expandafter\ifx\csname citenamefont\endcsname\relax
  \def\citenamefont#1{#1}\fi
\expandafter\ifx\csname url\endcsname\relax
  \def\url#1{\texttt{#1}}\fi
\expandafter\ifx\csname urlprefix\endcsname\relax\def\urlprefix{URL }\fi
\providecommand{\bibinfo}[2]{#2}
\providecommand{\eprint}[2][]{\url{#2}}

\bibitem[{\citenamefont{Timusk and
  Statt}(1999)}]{Timusk:Rep.Prog.Phys.:62(1999)61--122}
\bibinfo{author}{\bibfnamefont{T.}~\bibnamefont{Timusk}} \bibnamefont{and}
  \bibinfo{author}{\bibfnamefont{B.}~\bibnamefont{Statt}},
  \bibinfo{journal}{Rep. Prog. Phys.} \textbf{\bibinfo{volume}{62}},
  \bibinfo{pages}{61} (\bibinfo{year}{1999}).

\bibitem[{\citenamefont{Kanigel et~al.}(2006)\citenamefont{Kanigel, Norman,
  Randeria, Chatterjee, Souma, Kaminski, Fretwell, Rosenkranz, Shi, Sato
  et~al.}}]{Kanigel:NaturePhys.:2(2006)447--451}
\bibinfo{author}{\bibfnamefont{A.}~\bibnamefont{Kanigel}}  \bibnamefont{{\it et~al.}},
  \bibinfo{journal}{Nature Phys.} \textbf{\bibinfo{volume}{2}},
  \bibinfo{pages}{447} (\bibinfo{year}{2006}).

\bibitem[{\citenamefont{Hashimoto et~al.}(2010)\citenamefont{Hashimoto, He,
  Tanaka, Testaud, Meevasana, Moore, Lu, Yao, Yoshida, Eisaki
  et~al.}}]{Hashimoto:NatPhys:6(2010)414--418}
\bibinfo{author}{\bibfnamefont{M.}~\bibnamefont{Hashimoto}} \bibnamefont{et~al.},
\bibinfo{journal}{Nature Phys.}
  \textbf{\bibinfo{volume}{6}}, \bibinfo{pages}{414} (\bibinfo{year}{2010}).

\bibitem[{\citenamefont{Norman et~al.}(1998)\citenamefont{Norman, Ding,
  Randeria, Campuzano, Yokoya, Takeuchi, Takahashi, Mochiku, Kadowaki,
  Guptasarma et~al.}}]{Norman:Nature:392(1998)157--160}
\bibinfo{author}{\bibfnamefont{M.~R.} \bibnamefont{Norman}} \bibnamefont{et~al.},
\bibinfo{journal}{Nature}
  \textbf{\bibinfo{volume}{392}}, \bibinfo{pages}{157} (\bibinfo{year}{1998}).

\bibitem[{\citenamefont{Shen et~al.}(2005)\citenamefont{Shen, Ronning, Lu,
  Baumberger, Ingle, Lee, Meevasana, Kohsaka, Azuma, Takano
  et~al.}}]{Shen:Science:307(2005)901--904}
\bibinfo{author}{\bibfnamefont{K.~M.} \bibnamefont{Shen}} \bibnamefont{et~al.},
\bibinfo{journal}{Science}
  \textbf{\bibinfo{volume}{307}}, \bibinfo{pages}{901} (\bibinfo{year}{2005}).

\bibitem[{\citenamefont{Valla et~al.}(2006)\citenamefont{Valla, Fedorov, Lee,
  Davis, and Gu}}]{Valla:Science:314(2006)1914--1916}
\bibinfo{author}{\bibfnamefont{T.}~\bibnamefont{Valla}} \bibnamefont{et~al.},
\bibinfo{journal}{Science} \textbf{\bibinfo{volume}{314}},
  \bibinfo{pages}{1914} (\bibinfo{year}{2006}).

\bibitem[{\citenamefont{Doiron-Leyraud
  et~al.}(2007)\citenamefont{Doiron-Leyraud, Proust, LeBoeuf, Levallois,
  Bonnemaison, Liang, Bonn, Hardy, and
  Taillefer}}]{Doiron-Leyraud:Nature:447(2007)565--568}
\bibinfo{author}{\bibfnamefont{N.}~\bibnamefont{Doiron-Leyraud}} \bibnamefont{et~al.},
\bibinfo{journal}{Nature} \textbf{\bibinfo{volume}{447}},
  \bibinfo{pages}{565} (\bibinfo{year}{2007}).

\bibitem[{\citenamefont{Yelland et~al.}(2008)\citenamefont{Yelland, Singleton,
  Mielke, Harrison, Balakirev, Dabrowski, and
  Cooper}}]{Yelland:Phys.Rev.Lett.:100(2008)047003}
\bibinfo{author}{\bibfnamefont{E.~A.} \bibnamefont{Yelland}} \bibnamefont{et~al.},
\bibinfo{journal}{Phys. Rev. Lett.}
  \textbf{\bibinfo{volume}{100}}, \bibinfo{pages}{047003}
  (\bibinfo{year}{2008}).

\bibitem[{\citenamefont{Chakravarty et~al.}(2003)\citenamefont{Chakravarty,
  Nayak, and Tewari}}]{Chakravarty:Phys.Rev.B:68(2003)100504}
\bibinfo{author}{\bibfnamefont{S.}~\bibnamefont{Chakravarty}},
  \bibinfo{author}{\bibfnamefont{C.}~\bibnamefont{Nayak}}, \bibnamefont{and}
  \bibinfo{author}{\bibfnamefont{S.}~\bibnamefont{Tewari}},
  \bibinfo{journal}{Phys. Rev. B} \textbf{\bibinfo{volume}{68}},
  \bibinfo{pages}{100504} (\bibinfo{year}{2003}).

\bibitem[{\citenamefont{Yang et~al.}(2006)\citenamefont{Yang, Rice, and
  Zhang}}]{Yang:Phys.Rev.B:73(2006)174501}
\bibinfo{author}{\bibfnamefont{K.-Y.} \bibnamefont{Yang}},
  \bibinfo{author}{\bibfnamefont{T.~M.} \bibnamefont{Rice}}, \bibnamefont{and}
  \bibinfo{author}{\bibfnamefont{F.-C.} \bibnamefont{Zhang}},
  \bibinfo{journal}{Phys. Rev. B} \textbf{\bibinfo{volume}{73}},
  \bibinfo{pages}{174501} (\bibinfo{year}{2006}).

\bibitem[{\citenamefont{LeBoeuf et~al.}(2007)\citenamefont{LeBoeuf,
  Doiron-Leyraud, Levallois, Daou, Bonnemaison, Hussey, Balicas, Ramshaw,
  Liang, Bonn et~al.}}]{LeBoeuf:Nature:450(2007)533--536}
\bibinfo{author}{\bibfnamefont{D.}~\bibnamefont{LeBoeuf}} \bibnamefont{et~al.},
\bibinfo{journal}{Nature}
  \textbf{\bibinfo{volume}{450}}, \bibinfo{pages}{533} (\bibinfo{year}{2007}).

\bibitem[{\citenamefont{Yang et~al.}(2008)\citenamefont{Yang, Rameau, Johnson,
  Valla, Tsvelik, and Gu}}]{Yang:Nature:456(2008)77--80}
\bibinfo{author}{\bibfnamefont{H.-B.} \bibnamefont{Yang}} \bibnamefont{et~al.},
 \bibinfo{journal}{Nature} \textbf{\bibinfo{volume}{456}}, \bibinfo{pages}{77}
  (\bibinfo{year}{2008}).

\bibitem[{\citenamefont{Sebastian et~al.}(2008)\citenamefont{Sebastian,
  Harrison, Palm, Murphy, Mielke, Liang, Bonn, Hardy, and
  Lonzarich}}]{Sebastian:Nature:454(2008)200--203}
\bibinfo{author}{\bibfnamefont{S.~E.} \bibnamefont{Sebastian}} \bibnamefont{et~al.},
  \bibinfo{journal}{Nature} \textbf{\bibinfo{volume}{454}},
  \bibinfo{pages}{200} (\bibinfo{year}{2008}).

\bibitem[{\citenamefont{Meng et~al.}(2009)\citenamefont{Meng, Liu, Zhang, Zhao,
  Liu, Jia, Mu, Liu, Dong, Zhang et~al.}}]{Meng:Nature:462(2009)335--338}
\bibinfo{author}{\bibfnamefont{J.}~\bibnamefont{Meng}} \bibnamefont{et~al.},
\bibinfo{journal}{Nature}
  \textbf{\bibinfo{volume}{462}}, \bibinfo{pages}{335} (\bibinfo{year}{2009}).

\bibitem[{\citenamefont{Yang et~al.}(2010)\citenamefont{Yang, Ramaeu, Pan, Gu,
  Johnson, Claus, Hinks, and Kidd}}]{Yang:arXiv:1008.3121:(2010)}
\bibinfo{author}{\bibfnamefont{H.-B.} \bibnamefont{Yang}} \bibnamefont{et~al.},
  \bibinfo{journal}{arXiv:1008.3121}  (\bibinfo{year}{2010}).

\bibitem[{\citenamefont{Norman}(2010)}]{Norman:Physics:3(2010)86}
\bibinfo{author}{\bibfnamefont{M.~R.} \bibnamefont{Norman}},
  \bibinfo{journal}{Physics} \textbf{\bibinfo{volume}{3}}, \bibinfo{eid}{86}
  (\bibinfo{year}{2010}).

\bibitem[{\citenamefont{Subramanian et~al.}(1988)\citenamefont{Subramanian,
  Torardi, Calabrese, Gopalakrishan, Morrissey, Askew, Flippen, Chowdhry, and
  Sleight}}]{Subramanian:Science:239(1988)1015--1017}
\bibinfo{author}{\bibfnamefont{M.~A.} \bibnamefont{Subramanian}}  \bibnamefont{et~al.},
  \bibinfo{journal}{Science} \textbf{\bibinfo{volume}{239}},
  \bibinfo{pages}{1015} (\bibinfo{year}{1988}).

\bibitem[{\citenamefont{Torardi et~al.}(1988)\citenamefont{Torardi,
  Subramanian, Calabrese, Gopalakrishnan, McCarron, Morrissey, Askew, Flippen,
  Chowdhry, and Sleight}}]{Torardi:Phys.Rev.B:38(1988)225--231}
\bibinfo{author}{\bibfnamefont{C.~C.} \bibnamefont{Torardi}}  \bibnamefont{et~al.},
  \bibinfo{journal}{Phys. Rev. B} \textbf{\bibinfo{volume}{38}},
  \bibinfo{pages}{225} (\bibinfo{year}{1988}).

\bibitem[{\citenamefont{Mans et~al.}(2006)\citenamefont{Mans, Santoso, Huang,
  Siu, Tavaddod, Arpiainen, Lindroos, Berger, Strocov, Shi
  et~al.}}]{Mans:Phys.Rev.Lett.:96(2006)107007}
\bibinfo{author}{\bibfnamefont{A.}~\bibnamefont{Mans}}  \bibnamefont{et~al.},
  \bibinfo{journal}{Phys. Rev. Lett.} \textbf{\bibinfo{volume}{96}},
  \bibinfo{pages}{107007} (\bibinfo{year}{2006}).

\bibitem[{\citenamefont{Nakayama et~al.}(2006)\citenamefont{Nakayama, Sato,
  Dobashi, Terashima, Souma, Matsui, Takahashi, Campuzano, Kudo, Sasaki
  et~al.}}]{Nakayama:Phys.Rev.B:74(2006)054505}
\bibinfo{author}{\bibfnamefont{K.}~\bibnamefont{Nakayama}}  \bibnamefont{et~al.},
\bibinfo{journal}{Phys. Rev. B}
  \textbf{\bibinfo{volume}{74}}, \bibinfo{pages}{054505}
  (\bibinfo{year}{2006}).

\bibitem[{\citenamefont{Aebi et~al.}(1994)\citenamefont{Aebi, Osterwalder,
  Schwaller, Schlapbach, Shimoda, Mochiku, and
  Kadowaki}}]{Aebi:Phys.Rev.Lett.:72(1994)2757--2760}
\bibinfo{author}{\bibfnamefont{P.}~\bibnamefont{Aebi}}  \bibnamefont{et~al.},
  \bibinfo{journal}{Phys. Rev. Lett.} \textbf{\bibinfo{volume}{72}},
  \bibinfo{pages}{2757} (\bibinfo{year}{1994}).

\bibitem[{\citenamefont{Withers
  et~al.}(1988{\natexlab{a}})\citenamefont{Withers, Anderson, Hyde, Thompson,
  Wallenberg, FitzGerald, and Stewart}}]{Withers:J.Phys.C:21(1988)L417}
\bibinfo{author}{\bibfnamefont{R.~L.} \bibnamefont{Withers}}  \bibnamefont{et~al.},
\bibinfo{journal}{J. Phys. C}
  \textbf{\bibinfo{volume}{21}}, \bibinfo{pages}{L417}
  (\bibinfo{year}{1988}{\natexlab{a}}).

\bibitem[{\citenamefont{Withers
  et~al.}(1988{\natexlab{b}})\citenamefont{Withers, Thompson, Wallenberg,
  FitzGerald, Anderson, and Hyde}}]{Withers:J.Phys.C:21(1988)6067}
\bibinfo{author}{\bibfnamefont{R.~L.} \bibnamefont{Withers}}  \bibnamefont{et~al.},
  \bibinfo{journal}{J. Phys. C} \textbf{\bibinfo{volume}{21}},
  \bibinfo{pages}{6067} (\bibinfo{year}{1988}{\natexlab{b}}).

\bibitem[{\citenamefont{Yang et~al.}(1998)\citenamefont{Yang, Wen, Ni, Xiong,
  Chen, Dong, Wu, Qin, and Zhao}}]{Yang:PhysicaC:308(1998)294--300}
\bibinfo{author}{\bibfnamefont{W.~L.} \bibnamefont{Yang}}  \bibnamefont{et~al.},
  \bibinfo{journal}{Physica C} \textbf{\bibinfo{volume}{308}},
  \bibinfo{pages}{294} (\bibinfo{year}{1998}).

\bibitem{Dudy:J.Supercond.Nov.Magn.:22(2009)51--55}
\bibinfo{author}{\bibfnamefont{L.}~\bibnamefont{Dudy}}  \bibnamefont{et~al.},
\bibinfo{journal}{J. Supercond. Nov. Magn.}  \textbf{\bibinfo{volume}{22}},
\bibinfo{pages}{51} (\bibinfo{year}{2009}).

\bibitem[{\citenamefont{Zhang and
  Rice}(1988)}]{Zhang:Phys.Rev.B:37(1988)3759--3761}
\bibinfo{author}{\bibfnamefont{F.~C.} \bibnamefont{Zhang}} \bibnamefont{and}
  \bibinfo{author}{\bibfnamefont{T.~M.} \bibnamefont{Rice}},
  \bibinfo{journal}{Phys. Rev. B} \textbf{\bibinfo{volume}{37}},
  \bibinfo{pages}{3759} (\bibinfo{year}{1988}).

\bibitem[{\citenamefont{Chang et~al.}(2008)\citenamefont{Chang, Sassa,
  Guerrero, M̴nsson, Shi, Pailhs, Bendounan, Mottl, Claesson, Tjernberg
  et~al.}}]{Chang:NewJ.Phys.:10(2008)103016}
\bibinfo{author}{\bibfnamefont{J.}~\bibnamefont{Chang}}  \bibnamefont{et~al.},
\bibinfo{journal}{New J. Phys.}
  \textbf{\bibinfo{volume}{10}}, \bibinfo{pages}{103016}
  (\bibinfo{year}{2008}).

\bibitem[{\citenamefont{Hossain et~al.}(2008)\citenamefont{Hossain,
  Mottershead, Fournier, Bostwick, McChesney, Rotenberg, Liang, Hardy,
  Sawatzky, Elfimov et~al.}}]{Hossain:NaturePhys.:4(2008)527--531}
\bibinfo{author}{\bibfnamefont{M.~A.} \bibnamefont{Hossain}}  \bibnamefont{et~al.},
\bibinfo{journal}{Nature Phys.}
  \textbf{\bibinfo{volume}{4}}, \bibinfo{pages}{527} (\bibinfo{year}{2008}).

\end{thebibliography}

\end{document}